\documentclass[11pt,a4paper]{article}
\pdfoutput=1   
\usepackage[utf8]{inputenc}
\usepackage{tabularx}
\usepackage{lmodern}
\usepackage[outercaption]{sidecap}
\usepackage{xspace}
\usepackage{cancel}
\usepackage{enumerate}
\usepackage{booktabs}
\usepackage{mathtools}
\usepackage[section]{placeins}
\usepackage{alpha}
\hypersetup{%
   pdftitle    = {Low-mode deflation for twisted-mass and RHMC reweighting in lattice QCD},
   pdfauthor   = {S.~Kuberski},
   pdfkeywords = {}
}%

\begin{document}

\preprintno{%
MITP-23-021
\vfill
}

\title{%
Low-mode deflation for twisted-mass\\ and RHMC reweighting\\
in lattice QCD
}

\author{Simon Kuberski}

\address{simon.kuberski@uni-mainz.de\\
         Helmholtz Institute Mainz, Johannes Gutenberg University, 55099 Mainz, Germany,\\
         GSI Helmholtzzentrum für Schwerionenforschung, Planckstraße 1,
         64291 Darmstadt, Germany
        }

\begin{abstract}
We propose improved estimators to compute the reweighting factors which are 
needed for lattice QCD calculations that rely on twisted-mass reweighting 
for the light quark contribution and the Rational Hybrid Monte Carlo (RHMC)
algorithm for non-degenerate quark masses. This is the case for a number 
of modern large-scale simulations based on $\mathrm{O}(a)$ improved Wilson
fermions.
We find a significant reduction of uncertainties for the reweighting factors
at similar computational cost compared to the conventional estimation.
This leads to a significant increase in precision for phenomenologically
relevant observables with high correlation to the low eigenmodes of the 
Wilson-Dirac operator in the presence of exceptionally small eigenvalues.
Supplementary details regarding the spectral gap of the light quark Dirac
operator on the $2+1$ flavor large-volume ensembles explored in this study
can be found in an accompanying appendix.%
\\
\end{abstract}


\maketitle

\tableofcontents

\makeatletter
\g@addto@macro\bfseries{\boldmath}
\makeatother

\section{Introduction}
Lattice QCD is the only known framework for the calculation of ab initio 
predictions in the low-energy regime of Quantum Chromodynamics, the theory of the strong interactions in the Standard Model of particle physics. The field
has reached the stage where high precision lattice QCD calculations can have a
significant impact on phenomenologically relevant observables. 
Prime examples are the strong coupling constant $\alpha_{\rm s}$ and the 
hadronic contributions to the anomalous magnetic moment of the muon 
\cite{Aoyama:2020ynm,Colangelo:2022jxc}. For a recent review of lattice results
on flavor physics observables see 
\cite{FlavourLatticeAveragingGroupFLAG:2021npn}.

To control all sources of systematic uncertainties, simulations with resolutions
finer than $0.05\,$fm on lattices with spatial extents larger than $4\,$fm
and physical quark masses are desirable and by now state of the art.
The generation of gauge ensembles in lattice QCD
using the Wilson formulation and the Hybrid Monte Carlo (HMC) algorithm
\cite{Duane:1987de} is known to suffer from instabilities in 
large volumes and towards physical light quark masses 
\cite{DelDebbio:2005qa,DelDebbio:2007pz,DellaMorte:2004hs}.
This is caused by accidental near-zero modes of the light quark lattice Dirac
operator. Twisted-mass reweighting of the light quark determinant as 
introduced in  \cite{Luscher:2008tw, Luscher:2012av} greatly reduces
instabilities and sampling inefficiencies and thus allows for the simulation 
of QCD with Wilson fermions at the physical quark mass and in large volumes.

Twisted-mass reweighting is used in a number of modern large-scale simulations 
with $\mathrm{O}(a)$ improved Wilson fermions
\cite{Bruno:2014jqa,Francis:2019muy,Cuteri:2022oms,Hollwieser:2020qri}.
Among them, the CLS consortium has generated a large set of large-volume, $2+1$
flavor gauge ensembles  across six lattice spacings below $0.1\,$fm and a range 
of quark masses down to the physical values of the masses of the up and 
down quarks \cite{Bruno:2014jqa,Bali:2016umi,Mohler:2017wnb} 
using $\mathrm{O(a)}$ improved Wilson fermions 
\cite{Sheikholeslami:1985ij, Bulava:2013cta}  and the tree-level 
improved Lüscher-Weisz gauge action \cite{Luscher:1984xn}.%
\footnote{See Reference~\cite{RQCD:2022xux} for an overview of the current
set of ensembles. The ensembles included in this work are listed
in table \ref{t:enstab}.}%
Difficulties in accurately estimating the twisted-mass reweighting factor
have been encountered on a number of gauge configurations, as discussed in
Appendix G of \cite{RQCD:2022xux}, see also \cite{Bruno:2014lra}.
As pointed out in \cite{Bruno:2014jqa},
for observables that have a strong correlation 
with the reweighting factor, i.e., observables that are sensitive to the
low eigenmodes of the light quark Dirac operator, insufficient precision 
in the determination of the reweighting factor can lead to large 
fluctuations in the Monte Carlo history and correspondingly to 
significantly increased statistical errors. 
This behavior is particularly pronounced in the case of an anticorrelation of
reweighting factor and observable, since a large value of the observable and
a small reweighting factor must cancel to high precision on configurations
that are affected by strong fluctuations.

Traditionally, the calculation of the twisted-mass reweighting factor has been
performed via stochastic estimation. It was already anticipated in 
\cite{Luscher:2008tw, Luscher:2012av} that improved estimators might be 
needed for high-precision studies at near-physical light quark masses.
A viable solution to improve the estimation of reweighting factors with
high correlation with the low eigenmodes of the Dirac operator has been introduced in \cite{Hasenfratz:2008fg} in the context of quark mass
reweighting: It combines the exact computation of the reweighting factor in
the subspace of the lowest eigenmodes of the Dirac operator with a
stochastic estimation in the corresponding orthogonal subspace. 
Such an estimator improves the computation because the low modes contribute
significantly to the reweighting factor and its stochastic fluctuations.
In this work, we develop and apply low-mode deflation to the computation of 
twisted-mass reweighting factors. We find that the improved setup effectively
solves the accuracy problems in the computation with similar or even less cost
than the purely stochastic setup. 

Another reweighting factor has to be considered when the 
Rational Hybrid Monte Carlo (RHMC) algorithm
\cite{Kennedy:1998cu,Clark:2003na, Clark:2006fx}
is employed to simulate fermions with non-degenerate masses,
as it is the case in  state-of-the-art simulations, see for example \cite{Francis:2019muy,Cuteri:2022oms,Hollwieser:2020qri,Ishikawa:2018jee, Shintani:2019wai,Campos:2019kgw,RCstar:2022yjz}.
It corrects for the approximation of the inverse square root of the Dirac
operator in the RHMC. 
Whereas this reweighting factor is not expected to fluctuate strongly for
appropriately chosen parameters, exceptionally small eigenvalues of the 
strange quark Dirac operator can hinder the precise computation. 
We find that a low-mode deflated setup may be used in 
these cases to resolve all precision issues.

This work is structured as follows: We discuss the computation of 
twisted-mass reweighting factors in section \ref{s:defl_tm}, where we
first recapitulate the underlying regularization of the quark determinant 
and the stochastic estimation of the reweighting factor before  
incorporating exact low-mode deflation into the computation.
We then turn to the case of the RHMC reweighting factor in section 
\ref{s:RHMC}, where we again give an overview of the stochastic estimation
before introducing the application of low-mode deflation.
The newly developed estimators are presented in sections \ref{s:TM_dfl} and
\ref{s:RHMC_dfl}, respectively.
In section \ref{s:num}, we present the effect of low-mode deflation on the 
computation of reweighting factors for the case of $2+1$ flavor CLS ensembles.
We summarize the expected impact of our findings
on ongoing and future lattice calculations in section \ref{s:conclusions}.
Appendix~\ref{a:gap} gathers some information on the behavior of the 
spectral gap of the light quark Dirac operator for $\mathrm{O}(a)$ 
improved Wilson fermions obtained in the course of this work.

\section{Twisted-mass reweighting\label{s:defl_tm}}
Twisted-mass reweighting is employed to stabilize HMC simulations 
of lattice QCD with very light Wilson quarks. The fermion determinant
is split up into two factors and only one of them is taken into account
in the HMC, whereas the factor that includes the contribution of the
low eigenmodes of the Dirac operator is treated as reweighting factor.
Before we turn to the application of low-mode deflation to the determination
of the twisted-mass reweighting factor, we will briefly recapitulate the 
definition of the regulator and the stochastic estimation of the 
reweighting factor.

\subsection{Twisted-mass determinant reweighting \label{s:theo_tm_stoch}}
There exists some freedom in the exact definition of the twisted-mass regulator. 
For definiteness, we here choose to work with the setup that is employed 
on the $2+1$ flavor CLS ensembles \cite{Bruno:2014jqa}, where the
regularization is applied to the Schur complement of the asymmetric 
even-odd preconditioning of the hermitian Dirac operator 
\cite{DeGrand:1988vx,openQCD:dirac},
\begin{align}
\hat{Q}_{\rm l} 
= 
Q_\mathrm{l, ee}^{\phantom{-1}} - Q_\mathrm{l, eo}^{\phantom{-1}} Q_\mathrm{l, oo}^{-1}Q_\mathrm{l, oe}^{\phantom{-1}}\,,
\quad\text{with}\quad
\qquad Q_{\rm l}
= 
\gamma_5 D_\mathrm{l}\,,
\end{align}
with the light quark Wilson-Dirac operator $D_{\rm l}^{\phantom{\dagger}}$. 
The second of the two variants in \cite{Luscher:2008tw} 
(see Table 1 of that reference) is employed for the regularization of the 
light quark determinant, as it is less affected by fluctuations from the
ultraviolet part of the spectrum of the Dirac operator
\cite{Luscher:2012av}.
This amounts to the replacement
\begin{align}
\det(Q^2_{\rm l}) = 
\det(Q_{\rm l,oo}^2)\det (\hat{Q}_{\rm l}^2) 
\rightarrow 
\det(Q_{\rm l,oo}^2)
\det \frac{
	(\hat{Q}_{\rm l}^2 + \mu_0^2)^2
}{
	(\hat{Q}_{\rm l}^2 + 2\mu_0^2)
}
\end{align}
where $\mu_0 > 0$ is the infrared regulator.
The reweighting factor, needed to restore the target distribution, 
is defined via
\begin{align}
	W_{\rm l} = \det \frac{(\hat{Q}^2_{\rm l} + 2\mu_0^2)\hat{Q}^2_{\rm l}}{(\hat{Q}^2_{\rm l} + \mu_0^2)^2}\,,
\end{align}
and is computed on the gauge configurations that have been generated
with the regularized determinant.
When computing expectation values of primary observables $O$ that have been 
computed on ensembles with the modified action, the reweighting factor 
$W_{\rm l}$ is taken into account by the redefinition of the 
gauge expectation value via
\begin{align}
\langle O \rangle \rightarrow \frac{\langle O\, W_{\rm l}\rangle}{\langle W_{\rm l} \rangle}\,.
\end{align}

Frequency splitting, similar to that introduced by Hasenbusch \cite{Hasenbusch:2001ne,Hasenbusch:2002ai}, may be applied to the 
computation of $W_{\rm l}$ \cite{Hasenfratz:2008fg}. 
In this case, the interval between 0 and $\mu_0$ is split into $N_\mathrm{sp}$ smaller intervals with $\mu_0= \tilde{\mu}_0 > \tilde{\mu}_1 > \dots > \tilde{\mu}_{N_\mathrm{sp}} = 0$ and $W_{\rm l}$ can be factorized,
\begin{align}
	W_{\rm l}(\mu_0) = \prod_{i=1}^{N_\mathrm{sp}}\det R(\tilde{\mu}_{i-1},\tilde{\mu}_i)\,,\qquad R(\mu_1, \mu_2) = \frac{(\hat{Q}_{\rm l}^2 + \mu_2^2)^2(\hat{Q}_{\rm l}^2 + 2\mu_1^2)}{(\hat{Q}_{\rm l}^2 + \mu_1^2)^2(\hat{Q}_{\rm l}^2 + 2\mu_2^2)}\,.
\end{align}
The determinants entering this product can be computed stochastically from the
estimator
\begin{align}
	\langle \tilde{R}(\mu_1, \mu_2, N_\mathrm{r})\rangle_\eta =
	\frac{1}{N_\mathrm{r}} \sum_{i=1}^{N_\mathrm{r}} \exp\left\{-(\eta_i, (R^{-1}(\mu_1,\mu_2)-1)\eta_i)\right\}\,,
\end{align}
where $\eta_i$ are complex-valued Gaussian distributed random fields of unit variance, defined on the even lattice sites.
The determinant of $R$ is related to this stochastic estimator via
\begin{align}
	\det R(\mu_1, \mu_2) \propto \langle \tilde{R}(\mu_1, \mu_2, N_\mathrm{r})\rangle_\eta\,, \label{e:Rtm_stoch}
\end{align}
up to an irrelevant constant. The variance of $\tilde{R}$ can be computed from
\begin{align}
	\text{Var}(\tilde{R}) = \langle\tilde{R}^2 \rangle_\eta - \langle\tilde{R} \rangle_\eta^2 = \det \frac{R}{2-R} - (\det R)^2 \,. \label{e:stochvar}
\end{align}
and the relative variance increases significantly when $R$ deviates strongly
from unity (note that, with $\mu_1 > \mu_2$, the spectrum of $R(\mu_1, \mu_2)$
is contained in the interval ($\mu_2^2/\mu_1^2$, 1)).
In the context of \cite{Bruno:2014jqa}, only one set of stochastic estimators,
i.e., $N_{\rm sp} = 1$ is used in the computation of the reweighting factor.
The use of frequency splitting for the computation of twisted-mass reweighting
factors has been investigated in \cite{Bruno2016The} on several configurations,
where the relative uncertainty of the stochastic estimator was 
found to be above 100\%. A reduction of this uncertainty to the $\mathrm{O}(10\%)$ level could be achieved by the use of frequency 
splitting and a significantly increased computational effort.

\subsection{Deflated determinant reweighting \label{sec:defl_det_rw}}
Low-mode deflation may help to reduce significant statistical fluctuations of the stochastic estimator $\tilde{R}$ in the presence of exceptionally low eigenvalues of the Dirac operator. Following \cite{Hasenfratz:2008fg}, where low-mode deflation has been applied in the context of quark mass reweighting, we define a 
general reweighting factor $W$ 
based on an operator $\Omega$ via
\begin{align}
	W = \det(\Omega)^{-1}\,. \label{e:generalRWF}
\end{align}
Introducing the hermitian projection operators $P$ and $\bar{P}$ satisfying
\begin{align}
	P = P^\dagger\,,\quad 
	PP = P\,,\quad 
	\bar{P} = (\mathbf{1} - P) = \bar{P}\bar{P}\,,\quad 
	P\bar{P}=\bar{P}P = 0\,,
\end{align}
we can decompose $\Omega$ into a block structure in the two orthogonal 
subspaces of $P$ and $\bar{P}$ via
\begin{align}
	\Omega = 
	\begin{pmatrix}
	P\Omega P	&	P\Omega\bar{P} \\
	\bar{P}\Omega P	&	\bar{P}\Omega\bar{P} \\
	\end{pmatrix}
	= \begin{pmatrix}
	1 & 0 \\
	\bar{P}\Omega P (P\Omega P)^{-1} & 1
	\end{pmatrix}
	\begin{pmatrix}
	P\Omega P & 0\\
	0 & K
	\end{pmatrix}
	\begin{pmatrix}
	1 & (P\Omega P)^{-1} P\Omega\bar{P}\\
	0 & 1
	\end{pmatrix}\,,
	\label{e:Qsplit_matrix}
\end{align}
where the ordering is such that the subspace of $P$ comes first. With
\begin{align}
	K &= \bar{P}\Omega\bar{P} - \bar{P}\Omega P (P\Omega P)^{-1} P\Omega\bar{P}  \label{e:Qsplit_K}
\end{align}
we can split the determinant of $\Omega$ in the contributions of the two
subspaces via
\begin{align}
	\det (\Omega) = \det(P\Omega P)\det(K)\,. \label{e:detOmega_gen}
\end{align}
If we define the projectors $P$ from $N_\mathrm{L}$ exact (low) eigenmodes of $\Omega$, denoted by $v_i$,
\begin{align}
	P \equiv \sum_{i=1}^{N_\mathrm{L}} |v_i\rangle\langle v_i|\,,
\end{align}
eqs.~(\ref{e:Qsplit_matrix}--\ref{e:detOmega_gen}) simplify, since all mixed terms of $P$, $\Omega$ and $\bar{P}$ vanish, leaving us with
\begin{align}
	\det (\Omega) = \det(P\Omega P)\det(\bar{P}\Omega\bar{P})\,. \label{e:detOmega}
\end{align}
In the following, we will only consider the case where eigenmodes of $\Omega$,
determined to high precision, are employed for the deflation. 
Inexact deflation has not been implemented in this work; however, the framework
can be straightforwardly extended, as the general decomposition in
eq.~(\ref{e:Qsplit_matrix}) is independent of the specific definition of $P$.

We can assume that the properties of the deflated estimators outlined below
remain unchanged when low-precision eigenmodes are utilized.
The reduced cost for the determination of the eigenmodes is then 
contrasted with higher cost for the evaluation of the reweighting factor $W$.
While the first term in eq.~(\ref{e:detOmega}) can be straightforwardly
computed with negligible cost when $P$ is constructed from exact eigenmodes of
$\Omega$, additional computations must be performed in the case of inexact
eigenmodes. For the application in this work, this involves inversions of the
Dirac operator.
Additionally, the second term in eq.~(\ref{e:Qsplit_K}) does not vanish in this
case and must be evaluated explicitly.

\subsection{Deflated twisted-mass determinant reweighting \label{s:TM_dfl}}
In the case of twisted-mass reweighting, as introduced in section 
\ref{s:theo_tm_stoch}, we work with
\begin{align}
	\Omega = \frac{(\hat{Q}_{\rm l}^2 + 2\mu_0^2)\hat{Q}_{\rm l}^2}{(\hat{Q}_{\rm l}^2 + \mu_0^2)^2}
\end{align} 
or, if we consider frequency splitting,
\begin{align}
	\Omega = R(\mu_1, \mu_2)\,,
\end{align}
as operator that defines the reweighting factor in eq.~(\ref{e:generalRWF}).
Eigenmodes of the even-odd preconditioned, hermitian light quark Dirac 
operator $\hat{Q}_{\rm l}$ are therefore eigenmodes of $\Omega$ and may be used to construct the projector $P$ and thereby deflate $\Omega$ 
without mixed contributions of $P$, $\Omega$ and $\bar{P}$.
We can compute the first term on the right hand side of 
eq.~(\ref{e:detOmega}) from $N_{\rm L}$ eigenvalues $\lambda_i$ of
$\hat{Q}_{\rm l}$ via
\begin{align}
	\det (P\Omega P) = \prod_{i=1}^{N_\mathrm{L}} \frac{(\lambda_i^2+\mu_2^2)^2(\lambda_i^2+2 \mu_1^2)}{(\lambda_i^2+\mu_1^2)^2(\lambda_i^2+2\mu_2^2)}
	\underset{\mu_2=0}{\overset{\mu_1=\mu_0}{=}} 
	\prod_{i=1}^{N_\mathrm{L}} \frac{(\lambda_i^2 + 2\mu_0^2)\lambda_i^2}{(\lambda_i^2 + \mu_0^2)^2}\,, \label{e:detPQP}
\end{align}
using 
\begin{align}
	\hat{Q}_{\rm l} |v_i\rangle = \lambda_i |v_i\rangle \,,\qquad P|v_i\rangle = |v_i\rangle\,.
\end{align}
The second factor $\det(\bar{P}\Omega\bar{P})$ in eq.~(\ref{e:detOmega}) may be estimated stochastically, similar to the standard stochastic setup of eq.~(\ref{e:Rtm_stoch}). 
To compute the determinant in the subspace which is orthogonal to the space of the low modes, defined via $\bar{P}$, it is enough to project out the 
low modes.
The only change compared to eq.~(\ref{e:Rtm_stoch}) is the use of deflated random sources $\bar{P}\eta_i$ in the inversion.\footnote{This exact deflation of the lowest eigenmodes may speed up the solution of the Dirac
equation, if it does not already rely on (inexact) low-mode deflation}%
The stochastic estimators for $\det(\bar{P}\Omega\bar{P})$ are then defined by
\begin{align}
\langle\tilde{R}'(\mu_1, \mu_2, N_\mathrm{r})\rangle_\eta 
= \frac{1}{N_\mathrm{r}} \sum_{i=1}^{N_\mathrm{r}} \exp\left\{-(\bar{P}\eta_i, (R^{-1}(\mu_1,\mu_2)-1)\bar{P}\eta_i)\right\}\,,
\end{align}
and we can compute the estimator for $W_{\rm l}$,
\begin{align}
	\tilde{W}_{\rm l} &\propto \prod_{i=1}^{N_\mathrm{sp}} \det(PR(\tilde{\mu}_{i-1}, \tilde{\mu}_i)P) \cdot \langle \tilde{R}'(\tilde{\mu}_{i-1}, \tilde{\mu}_i, N_\mathrm{r})\rangle_\eta \label{e:Wev_gen}\\
	& {\overset{N_\mathrm{sp}=1}{=}} 
	\det(PR(\mu_0, 0)P) \cdot \langle \tilde{R}'(\mu_0, 0, N_\mathrm{r})\rangle_\eta\,, \label{e:Wev}
\end{align}
up to an irrelevant constant where, in the second line, we have taken the 
limit of one single factor in the frequency splitting. The variance of the
stochastic estimator $\tilde{R}'$, according to eq.~(\ref{e:stochvar}), 
is now solely based on the high-mode contribution to the reweighting factor.
\section{RHMC reweighting\label{s:RHMC}}
The use of the rational approximation when including quarks in the simulation
that do not come in mass degenerate pairs requires the inclusion of 
an additional reweighting factor to correct for the small deviation from 
the target action.
In the context of $2+1$ and $2+1+1$ flavor simulations with Wilson fermions, 
this concerns the strange and charm quark components. If non-degenerate
up and down quark masses are simulated, the RHMC algorithm may be employed
for all quark flavors \cite{Campos:2019kgw,RCstar:2022yjz}.
The idea of deflated reweighting factors, as discussed in section
\ref{sec:defl_det_rw}, may also be applied to the RHMC reweighting factor. 
We begin this section with a brief summary of RHMC reweighting \cite{Kennedy:1998cu,Clark:2003na, Clark:2006fx}, 
based on the notation used in \cite{openQCD:rhmc,Campos:2019kgw},
before including low-mode deflation.

\subsection{The RHMC quark determinant}
We consider the quark determinant for flavor $q$ in its even-odd preconditioned
form,
\begin{align}
	\det D_{q} = \det(D_{q, \mathrm{oo}}) \det(\hat{D}_{q})\,,
\end{align}
for the application in the RHMC algorithm.
The odd-odd part of the Dirac operator, $D_{q, \mathrm{oo}}$, 
can be directly included in the molecular-dynamics evolution. 
Therefore, we only need to handle the even-odd preconditioned operator, 
similar to the twisted-mass case. We write
\begin{align}
	\det \hat{D}_{q} = W_q \det(R^{-1}_q)\,, \label{e:detDhat_s}
\end{align}
where $R_q$ is an operator that approximates $(\hat{D}_q^\dagger\hat D_q^{\phantom{\dagger}})^{-1/2}$ and the reweighting factor is defined from
\begin{align}
	W_q = \det(\hat{D}_q R_q)\,. \label{e:Wsdef}
\end{align}
To approximate the inverse square root function $1/\sqrt{y}$ 
in the range $\epsilon \leq y \leq 1$, the Zolotarev function
\cite{achieser2013theory} with known coefficients $A, a_i$,
\begin{align}
R_{n, \epsilon}(y) = A \frac{(y+a_1)(y+a_3)\dots(y+a_{2n-1})}{(y-a_2)(y-a_4)\dots(y-a_{2n})}\,, \label{e:rhmc_r}
\end{align}
of degree $[n,n]$ may be used.
In simulations based on \texttt{openQCD}\cite{openQCD:rhmc}, as the ones
described in \cite{Bruno:2014jqa}, 
the representation of eq.~(\ref{e:rhmc_r}) is employed to approximate 
the operator $R_q$ of eq.~(\ref{e:detDhat_s}) via
\begin{align}
	R_q = r_\mathrm{b}^{-1} R_{n, \epsilon}(r_\mathrm{b}^{-2} \hat{D}_q^\dagger\hat D^{\phantom{\dagger}}_q)\,,\qquad \epsilon = (r_\mathrm{a} / r_\mathrm{b})^2\,.
\end{align}
The scalar factors $r_{\rm a}$, $r_{\rm b}$ define the approximation region
$[r_\mathrm{a}, r_\mathrm{b}]$, which has to cover the spectral range 
of the operator 
$(\hat{D}_q^\dagger\hat D^{\phantom{\dagger}}_q)^{1/2} = \left|\gamma_5 \hat{D}_q\right|$
in order to achieve a good approximation, resulting in a small deviation from
the target distribution. Due to the rather large mass of the strange quark,
$\hat{Q}_{\rm s}$ was assumed to have a safe spectral gap, although chiral symmetry is broken for Wilson fermions at finite lattice spacing. In \cite{Mohler:2020txx} it was shown, that this assumption is problematic at coarse lattice spacing, where configurations with negative strange quark determinants can be found. 
Another reweighting factor has been introduced to correct for the neglected sign and to restore the proper Boltzmann weight in the path integral.

The Zolotarev rational function can be split into several factors to achieve a
frequency splitting of the quark determinant.
We do not discuss this modification in the following, since the application of exact low-mode deflation can be trivially extended to the modified estimator.

\subsection{RHMC reweighting \label{s:RHMC_stoch}}
The reweighting factor $W_q$ can be determined by stochastic estimation. 
This can be done by writing \cite{openQCD:rhmc},
\begin{align}
	W_q = \det(\hat{D}_q R_q) = \det([\hat{D}^\dagger_q \hat{D}_q^{\phantom{\dagger}} R_q^2]^{1/2}) = \det([1+Z_q]^{1/2})\,,\qquad Z_q\equiv \hat{D}^\dagger_q \hat{D}_q^{\phantom{\dagger}} R_q^2 - 1\,,
\end{align}
with the newly introduced operator $Z_q$. 
The stochastic estimation is then performed by computing
\begin{align}
	\langle \tilde{W}_{q} \rangle_\eta 
	= \frac{1}{N_\mathrm{r}} \sum_{j=1}^{N_\mathrm{r}} \exp\{- (\eta_j, [(1+Z_q)^{-1/2} -1] \eta_j)\}\,,
\end{align}
using $N_\mathrm{r}$ random quark fields $\eta_j(x)$ with normal distribution, 
defined on the even lattice sites. The use of $Z_q$ is profitable here, because one can use the series
\begin{align}
	(1+ Z_q)^{-1/2} = 1 - \frac{1}{2} Z_q + \frac{3}{8}Z_q^2 - \frac{15}{16} Z_q^3 + \dots
\end{align}
to evaluate the square root in the stochastic estimator via
\begin{align}
	(\eta_j, [(1+Z_q)^{-1/2} -1] \eta_j) = -\frac{1}{2} (\eta_j, Z_q \eta_j) + \frac{3}{8} (\eta_j, Z_q^2 \eta_j) - \dots\,, \label{e:RHMC_Z_Taylor}
\end{align}
provided that the series expansion converges rapidly (i.e., the approximation is good). In \cite{openQCD:rhmc} it is shown that this is the case, if $[r_\mathrm{a}, r_\mathrm{b}]$ covers the spectral range of $\hat{Q}_q$.
Furthermore, arguments are given that in this case the fluctuations derived from the random sources can be assumed to be subleading with respect to the fluctuations from the gauge field and therefore $N_{\rm r}=1$ can be chosen.

Simulations of QCD with $2+1$ flavors of Wilson quarks using the RHMC algorithm
have been studied in \cite{Mohler:2020txx}. It was found that regions
in the configuration space where the strange quark Wilson-Dirac operator has 
negative eigenvalues are not negligible, contrary to standard assumptions. 
Since only the modulus of the fermion determinant is taken into account in 
the above procedure, it is not affected
by negative signs. However, these signs have to be taken into account as an
additional reweighting factor to restore the correct probability density in 
the path integral. Strategies for computing this sign are outlined in
\cite{Mohler:2020txx,RCstar:2022yjz}.%
\footnote{We note in passing that a small portion of the spectrum of 
the non-hermitian Dirac operator $\hat{D}_{\rm s}$ around the origin can be
computed using the \texttt{SLEPc} library \cite{slepc-toms} in combination
with \texttt{openQCD} to examine the signs of the lowest eigenvalues.
We have found that such a calculation confirms the results of
\cite{Mohler:2020txx}, where the mass dependence of the eigenvalues of
$\hat{Q}_{q}$ has been investigated instead.}

As explained in \cite{Mohler:2020txx}, changing the number of negative
eigenvalues of the strange quark Dirac operator by one unit can only be
achieved by passing through configurations with a zero eigenvalue of 
$D_{\rm s}$, which have a vanishing probability weight in the path integral. 
The use of the rational approximation in the RHMC leads to a
regularization of this infinite barrier between positive and negative
eigenvalues. 
In \cite{Mohler:2020txx} it is proposed to use this feature actively to allow 
the algorithm to tunnel between the two sectors and thus avoid possible
ergodicity problems.

The appearance of configurations with exceptionally small eigenvalues of
$\hat{Q}_{\rm s}$, which are much smaller than $r_{\rm a}$, cannot be
excluded in this case. The associated reweighting factor is intended to 
correct for such fluctuations. However, if the modulus of a near-zero
eigenvalue is too small, the computation of $W_{\rm s}$ by means of the
Taylor expansion in eq.~(\ref{e:RHMC_Z_Taylor}) fails because the
series does not converge within a tolerable number of terms.

\subsection{Deflated RHMC reweighting \label{s:RHMC_dfl}}
Low-mode deflation of the RHMC reweighting factor can resolve the issue
described above. We start with the factorization of the determinant of an
operator $\Omega$ in eq.~(\ref{e:detOmega_gen}). In the case that the projector
$P$ is built out of eigenmodes of $\Omega$ we may use eq.~(\ref{e:detOmega}).
In the computation of $W_q$, we employ
\begin{align}
	\Omega = (\hat{D}^\dagger_q \hat{D}_q^{\phantom{\dagger}} R_q^2)^{1/2}
\end{align}
and therefore eigenmodes of $\hat{Q}_q = \gamma_5 \hat{D}_q$ are
eigenmodes of $\Omega$. Using 
\begin{align}
\hat{Q}_{q} |v_i\rangle = \lambda_i |v_i\rangle\,, \qquad P \equiv \sum_{i=1}^{N_\mathrm{L}} |v_i\rangle\langle v_i|\,,
\end{align}
where $v_i$ are the eigenmodes of the hermitian Dirac operator $\hat{Q}_q$, we can compute 
\begin{align}
	\det(P\Omega P) = \prod_{i=1}^{N_\mathrm{L}} \sqrt{y_i} A \frac{(y_i+a_1)(y_i+a_3)\dots(y_i+a_{2n-1})}{(y_i-a_2)(y_i-a_4)\dots(y_i-a_{2n})}\,,\quad \text{with}\quad y_i \equiv \frac{\lambda_i^2}{r_\mathrm{b}^2}
	\label{e:Ws_POP}
\end{align}
from $N_{\rm L}$ real-valued eigenvalues $\lambda_i$ of $\hat{Q}_q$ and the known
coefficients $A$ and $a_j$ of the Zolotarev function.

The computation of $\det(\bar{P}\Omega\bar{P})$ can be done by stochastic estimation in the subspace that is orthogonal to the space of the low modes via
the estimator 
\begin{align}
	\langle \tilde{W}_{q}^{\prime} \rangle_\eta = \frac{1}{N_\mathrm{r}} \sum_{j=1}^{N_\mathrm{r}} \exp\{- (\bar{P}\eta_j, [(1+Z)^{-1/2} -1] \bar{P}\eta_j)\}\,,
\end{align}
where the issues related to the convergence of the series expansion are eliminated
since all eigenvalues of $\bar{P} \hat{Q}_q$ are greater than $r_{\rm a}$.
The only change in existing implementations of the stochastic estimation is the projection of the random sources using $\bar{P}$ and all other refinements,
such as the frequency splitting, remain unchanged.

\section{Numerical results \label{s:num}}
We have implemented the computation of deflated reweighting factors,
as described in the previous sections, by modifying existing routines of the \texttt{openQCD} package. 
The computation of a small number of low eigenmodes 
of $\hat{Q}_q$ is performed
using the \texttt{PRIMME} package \cite{PRIMME,svds_software}. 
In this section, we present the properties of the corresponding
reweighting factors on a number of $2+1$ flavor CLS ensembles. 
For an overview of all gauge ensembles considered in this work see 
table~\ref{t:enstab}. They cover a wide range of lattice spacings
and quark masses, including the physical ones.

\subsection{Application of deflated twisted-mass reweighting \label{s:num_tm}}
Stochastic evaluation of twisted-mass reweighting factors can become
problematic on configurations where the lowest eigenmode has an exceptionally
small eigenvalue, compared to the ensemble average and compared to the 
parameter $\mu_0$ that has been used in the course of the simulation. 
This is precisely the situation where we expect low-mode deflation to 
be most effective. To illustrate this assumption, we examine the relative
standard deviation of the stochastic estimators, configuration by configuration,
and compare the pure stochastic setup with a deflated computation.

We perform the comparison on the three ensembles N451, D450 and D452 with 
approximate pion masses 289\,MeV, 219\,MeV and 156\,MeV at a bare inverse
gauge coupling of $\beta=3.46$, corresponding to a relatively
coarse lattice spacing of $\approx 0.075\,$fm. 
In this regime, especially towards physical quark masses, we observe 
an increased number of configurations with an exceptionally small spectral 
gap of the light quark Dirac operator, leading to small twisted-mass 
reweighting factors. The Dirac operator does not have a solid spectral
gap here, since chiral symmetry is broken for Wilson fermions at finite 
lattice spacing.
In the stochastic setup, 24 random sources were used
to determine $W_{\rm l}$ on N451 and D450 and 36 sources on D452, 
respectively. For the computation of the deflated reweighting factors, 
we employ four low modes of $\hat{Q}_{\rm l}$ and reduce the number of 
random sources by a factor of two, compared to the stochastic setup.

\begin{figure}
	\centering
	\includegraphics[width=\textwidth]{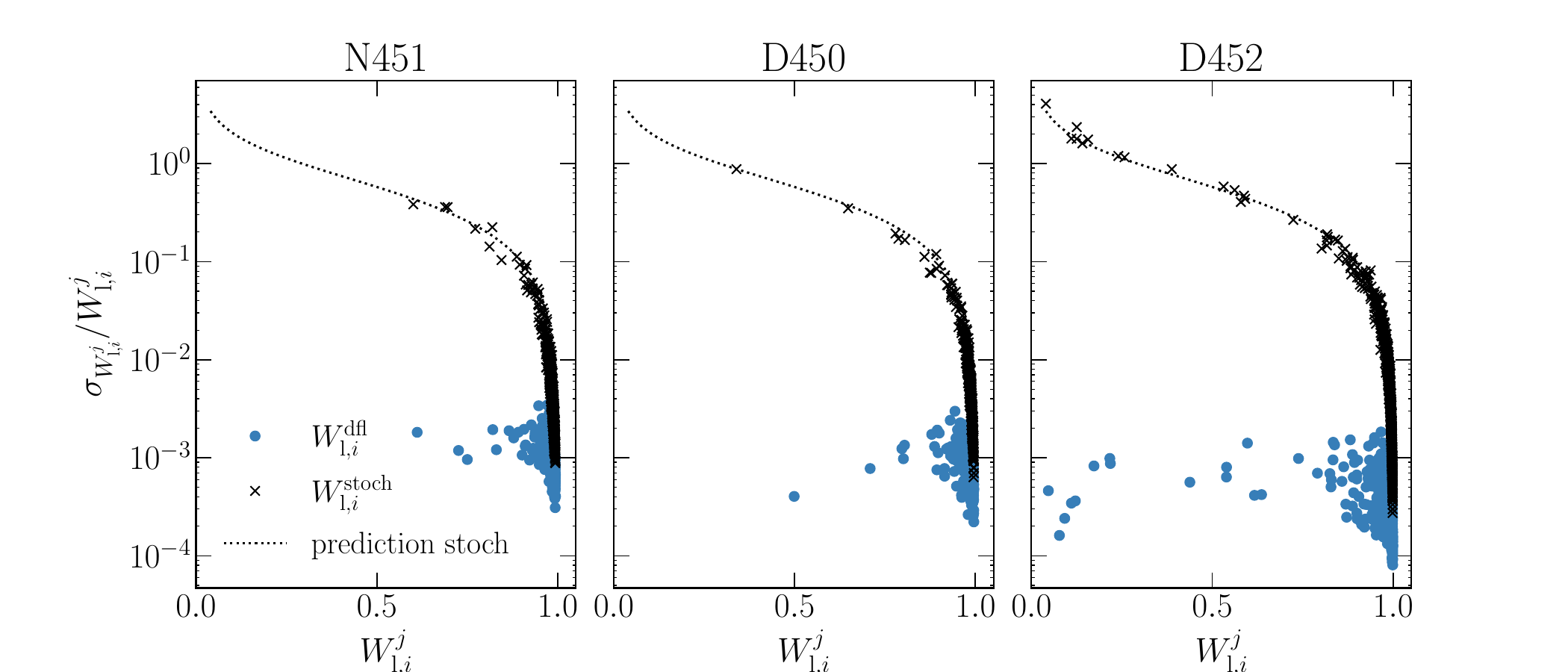}
	\caption{
		Dependence of the relative standard deviation of the stochastic estimator $W^\mathrm{stoch}_{\rm l}$ and the deflated estimator $W^\mathrm{dfl}_{\rm l}$ on the value of the estimator on the ensembles N451, D450 and D452 at $\beta=3.46$.
		The dotted line shows the prediction for the standard deviation 
		of $W^\mathrm{stoch}_{\rm l}$ based on eq.~(\ref{e:stochvar}). 
		Four low modes have been used in the deflated estimator.
		\label{f:varW346}}
\end{figure}

Fig.~\ref{f:varW346} shows the relative standard deviation of the stochastic estimator $W^\mathrm{stoch}_{{\rm l}, i}$ and of the stochastic contribution
to the deflated estimator $W^\mathrm{dfl}_{{\rm l}, i}$, depending on the
central value of the estimator on each configuration $i$ of the three Monte
Carlo chains. 
We also show the prediction of this dependence for 
$W^\mathrm{stoch}_{{\rm l}, i}$ based on eq.~(\ref{e:stochvar}). It is clearly visible that the data follow the prediction, such that a drastic increase in the relative standard deviation is observed as the central value of the reweighting factor decreases. The relative standard deviation of the deflated
estimator on the other hand is independent of the value of the reweighting factor, since only a small deviation from unity has to be estimated 
stochastically. Thus, by using low-mode deflation, we obtain a
reduction of the relative uncertainty by several orders of magnitude
in the regime of small reweighting factors. The gain is therefore
maximized on those configurations where we need to determine the reweighting
factor precisely because it is significantly different from one.

\begin{figure}[th]
	\centering
	\includegraphics[width=\textwidth]{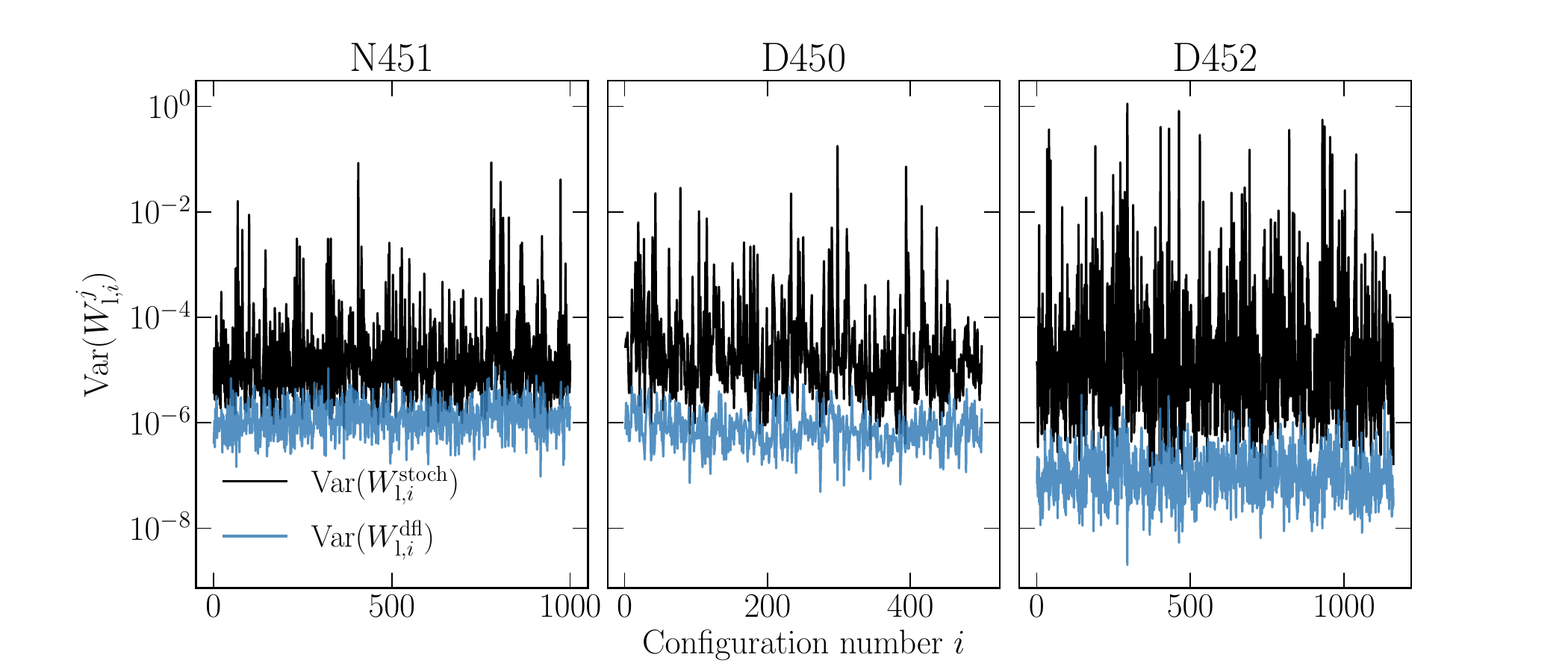}
	\caption{
		Monte Carlo history of the stochastic variances of the estimator
		$W^\mathrm{stoch}_{\rm l}$ and of the  deflated estimator $W^\mathrm{dfl}_{\rm l}$ on the ensembles N451, D450 and D452 at $\beta=3.46$.
		\label{f:varMC346}}
\end{figure}

At the same time, we also observe an overall reduction of the variance, as
depicted in Figure~\ref{f:varMC346}, where we show the Monte Carlo history
of the variance of the two estimators on the three ensembles under investigation.
Furthermore, the spread of the distribution decreases and no large spikes 
in the variance are present, when deflation is applied. 

\begin{figure}[ht]
	\centering
	\includegraphics[width=.85\textwidth]{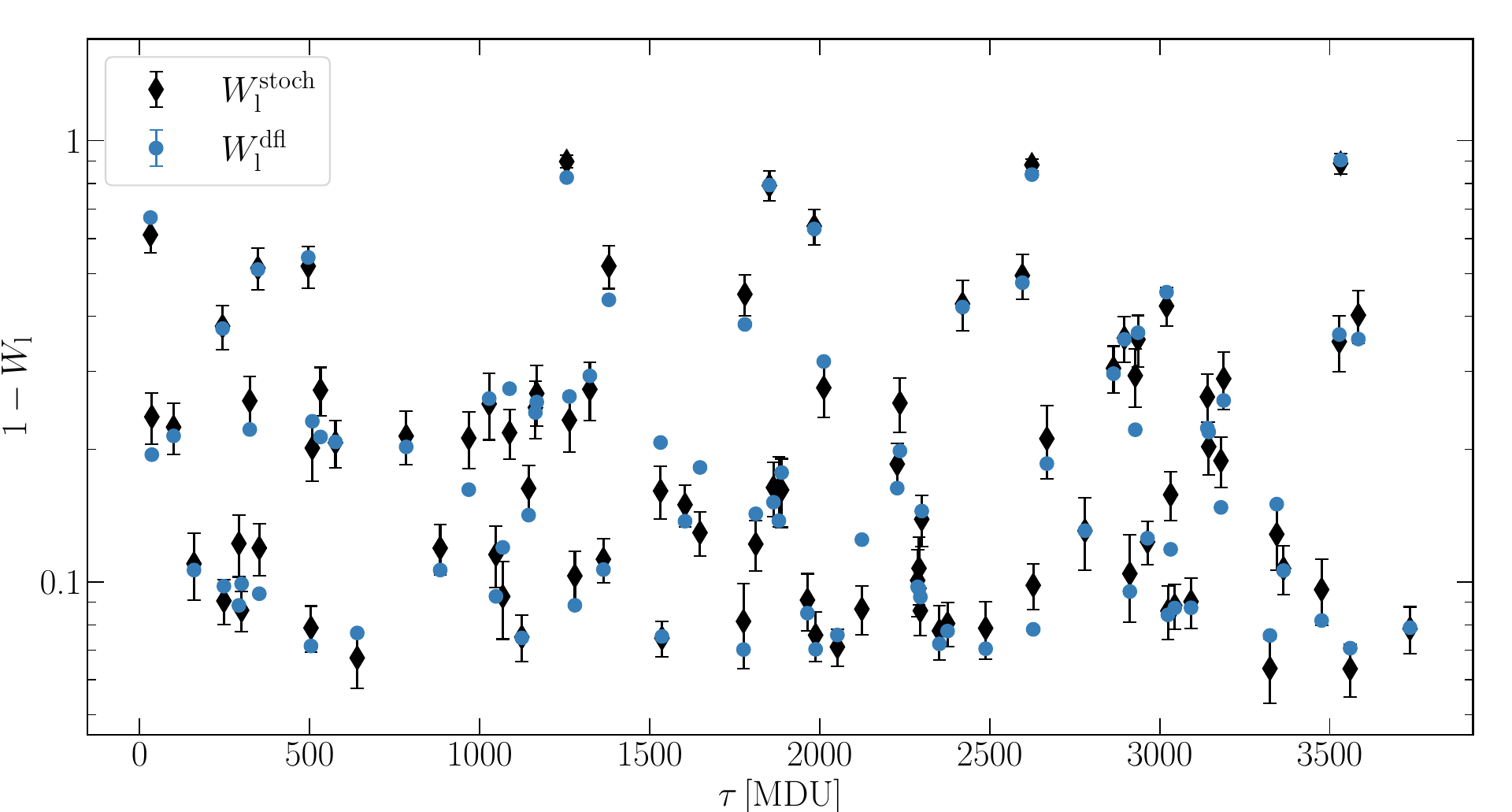}
	\caption{Deviation from one for the two sets of twisted-mass reweighting factors on ensemble E250r001 at close-to-physical pion mass. For clarity we show only factors smaller than 0.93. The error bars are derived from the distribution of the stochastic estimates on each configuration. The uncertainties on $W_{\rm l}^{\rm dfl}$ are smaller than the symbol size. \label{f:E250_compare}}
\end{figure}

In Figure~\ref{f:E250_compare} we directly compare twisted-mass reweighting factors determined in the stochastic setup with 24 sources
on ensemble E250 ($m_\pi=132\,$MeV, $a\approx 0.064\,$fm) \cite{Mohler:2017wnb}
with deflated ones, using four low modes and 12 stochastic sources. 
For better visibility we plot the deviation of the reweighting factors 
from unity and show only the factors on configurations $i$ where 
$W_{{\rm l}, i}^{\rm dfl} < 0.93$. 
Despite the approximate overall agreement, we can observe significant
deviations between the two sets of estimators for a number of configurations.
Note that the error estimates that are shown in the figure themselves carry 
a significant uncertainty since they have been computed from only 24 samples 
per configuration and assume a Gaussian distribution of the stochastic samples
that is not necessarily found in the data, see also the discussion 
in Appendix G of \cite{RQCD:2022xux}.
Such strong deviations do not affect the correctness of the reweighting
procedure but may lead to increased fluctuations in reweighted observables.

The cost of calculating the deflated
reweighting factor in the chosen setup is similar to the
purely stochastic computation: The calculation of the four lowest eigenmodes
of $\hat{Q}_{\rm l}$ with \texttt{PRIMME}, preconditioned by low precision
solves of the deflated solver of the \texttt{openQCD} package, is as expensive
as the computation of the reweighting factor on 12 random sources. 
The cost of the deflated computation can be further reduced,
considering that we have never observed a case where two lowest 
eigenmodes have an exceptionally small eigenvalue. Therefore, any significant
deviation of $W_{\rm l}$ from one is due to the lowest eigenmode and deflation
of a single mode is sufficient to achieve high precision. 
At the same time, also the number of random sources may be further reduced,
both compared to the stochastic setup and also compared to the setup that was
chosen here, to achieve sufficient precision.

\begin{figure}[th]
	\centering
	\includegraphics[width=.90\textwidth]{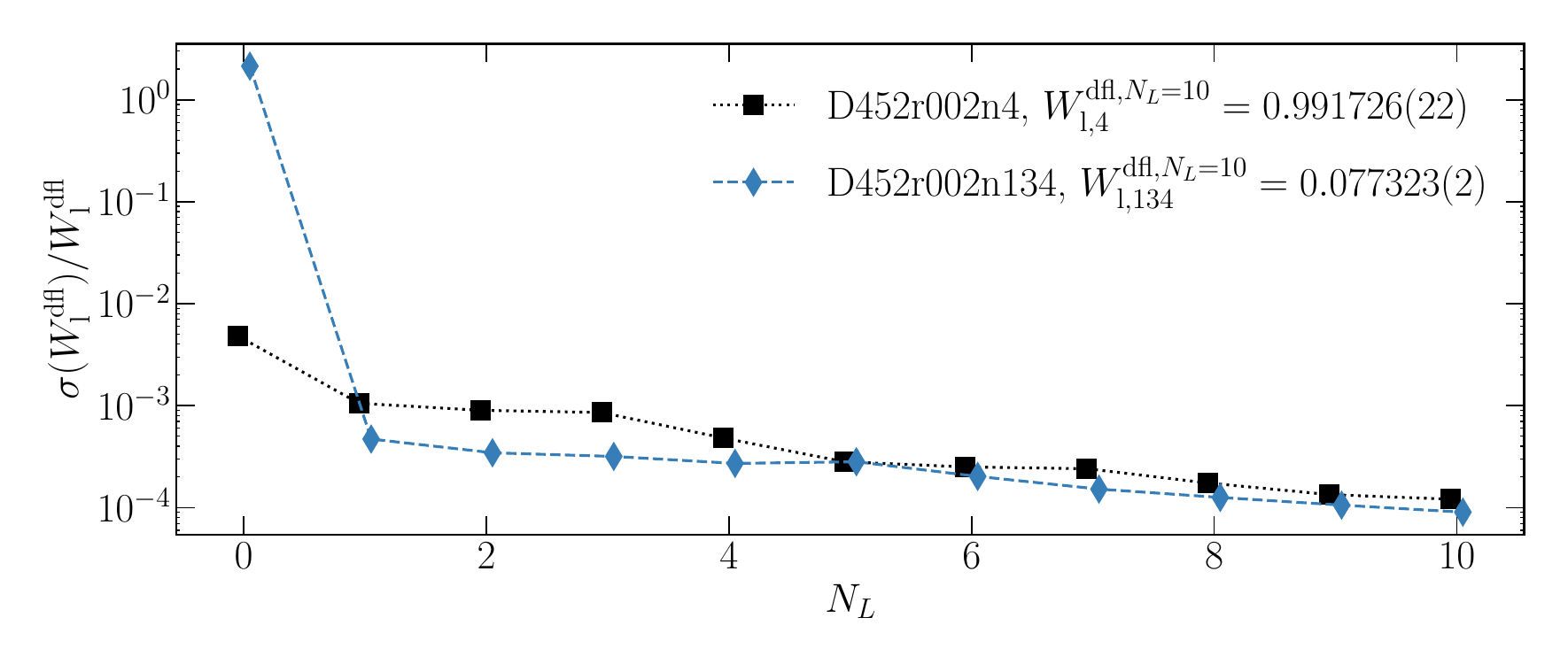}
	\caption{Relative standard deviation of the stochastic estimator 
	for $W_{\rm l}$ depending on the number of low modes $N_{\rm L}$ that is projected out, starting with $N_L=0$, i.e., no deflation, for D452r002n4 and D452r002n134. Lines are drawn to guide the eye. \label{f:D452_varNL}}
\end{figure}

In Fig.~\ref{f:D452_varNL} we show the dependence of the variance of the deflated stochastic estimator on the number of low modes $N_L$, starting with $N_L=0$, which is equivalent to the traditional stochastic estimator. 32 random sources are used for the stochastic estimation and for the estimation of the
variance.
We illustrate the behavior on two configurations of the D452 ensemble. 
$W_{\rm l}$ on configuration D452r002n4 is close to unity. Accordingly, the
estimator is well determined without deflation. However, increasing the number
of low modes significantly reduces the error of the stochastic part, about an order of magnitude when increasing $N_L$ from one to ten. For D452r002n134, where the reweighting factor is significantly different from one, 
deflation of the lowest mode has a significant impact on the relative error. 
For a larger number of modes, the behavior is similar to D452r002n4. 
The reduction in relative error below the per mil level 
that can be achieved by increasing the number of low modes is not necessary 
in practical applications. However, to achieve a target precision with minimal
cost, it may be beneficial to use more low modes and fewer stochastic sources.

The impact of the improved estimator for the reweighting factor on 
the expectation value of an observable depends strongly
on the observable in question. If its correlation with the low modes
of the Dirac operator is large, an observable will be strongly influenced by
exceptionally low eigenvalues.
One particularly affected observable is the pion decay constant $f_\pi$
which is often used to set the scale in lattice QCD computations, e.g.\ in \cite{Bruno:2016plf, Strassberger:2021tsu}, and may enter the determination 
of the hadronic vacuum polarization contribution to the anomalous magnetic
moment of the muon via $f_\pi$-rescaling \cite{Gerardin:2019rua, Ce:2022kxy}.

A significant number of exceptionally small eigenvalues of the light Dirac
operator can be found on the ensemble D150 
($m_\pi=132\,$MeV, $a\approx 0.085\,$fm). 
The bare decay constant $af_\pi$, computed as outlined in 
\cite{Ce:2022kxy, Ce:2022eix} from stochastically estimated correlation
functions and reweighting factors (using four stochastic time slice sources 
per configuration for the correlation functions%
\footnote{To be precise, we use four spin-diluted stochastic time slice sources
\cite{ETM:2008zte} per configuration, where the time slice is randomly chosen
for each source to make use of the translational invariance in the temporal
direction.}
and 32 sources for the 
twisted-mass reweighting factor) is determined to be $0.06808(155)$ on 
this ensemble.
When a deflated twisted-mass reweighting factor is used, the decay 
constant is evaluated to be $0.06837(74)$. Thus, we find a reduction of the
error by a factor of two, simply by using a more precise reweighting factor. 
If a robust estimator for the standard deviation is utilized, such as the
median absolute deviation \cite{doi:10.1080/01621459.1974.10482962}, 
we observe that the resulting error estimate, $5.5\cdot10^{-4}$, 
remains invariant under the exchange of reweighting factors within the 
error of the error.
This implies that outliers within the Monte Carlo distribution significantly
influence the variance estimation. These outliers, found on
configurations with exceptionally small eigenvalues of $\hat{Q}_{\rm l}$, are
effectively mitigated when employing more precise deflated reweighting factors.

The interplay of observable and reweighting factor also depends on the 
estimator that is used for the observable. It has been found that correlation 
functions constructed from stochastic time slice sources, as used in this 
exemplary case, have significantly fewer outliers in the MC
history than those built from point sources, such as those
used in hadron structure calculations. However, in both cases
potentially large local fluctuations of the eigenmodes cannot be completely
eliminated by the reweighting factor, which is based on the full four-volume
of the box.
Correlation functions determined by low-mode averaging (LMA)
\cite{Giusti:2004yp, DeGrand:2004qw, Giusti:2005sx}, on the other hand, include
a full volume average of the contribution of the lowest modes of the Dirac
operator. Therefore, we expect that large fluctuations due to the 
lowest mode are exactly canceled in combination with deflated reweighting
factors.

\begin{figure}[ht]
	\centering
	\includegraphics[width=.95\textwidth]{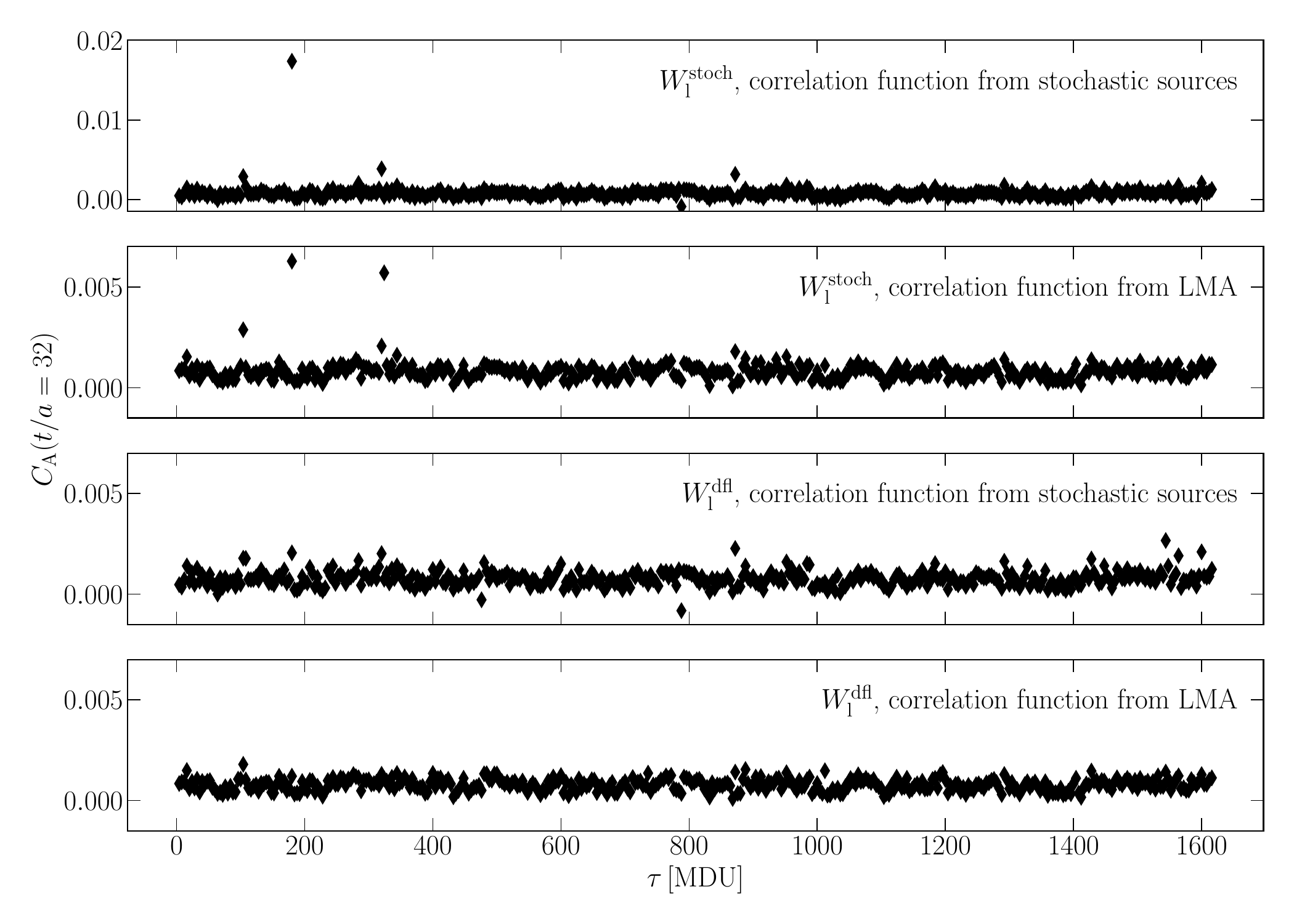}
	\caption{%
	Monte Carlo history of the light quark two-point correlation
	function based on pseudoscalar and axial density
    at source-sink separation $T/4$ 
	on ensemble D150 for four combinations of
	reweighting factor and correlation function. 
	Note the different scale in the first panel.
	\label{f:D150_fpi}}
\end{figure}

We illustrate the removal of outliers in the Monte Carlo history by using
improved estimators for the case of the light quark two-point function 
based on pseudoscalar and axial densities 
(defined as $C_A$ in \cite{Ce:2022kxy}) in figure~\ref{f:D150_fpi}. 
This correlation function enters the computation of $af_\pi$. 
We observe a significant reduction in the size and the number of 
spikes when low-mode deflation is used for the reweighting factor and a
further reduction when LMA is used to compute the correlation function. 
Note that a larger scale is used to depict the fluctuations
for the combination of the stochastically estimated reweighting factor 
and correlation function in the top panel.
The variance derived from the mean absolute deviation is the same in 
all four cases considered here.

\subsection{Application of deflated RHMC reweighting \label{s:num_rhmc}}

\begin{figure}[th]
	\centering
	\includegraphics[width=.90\textwidth]{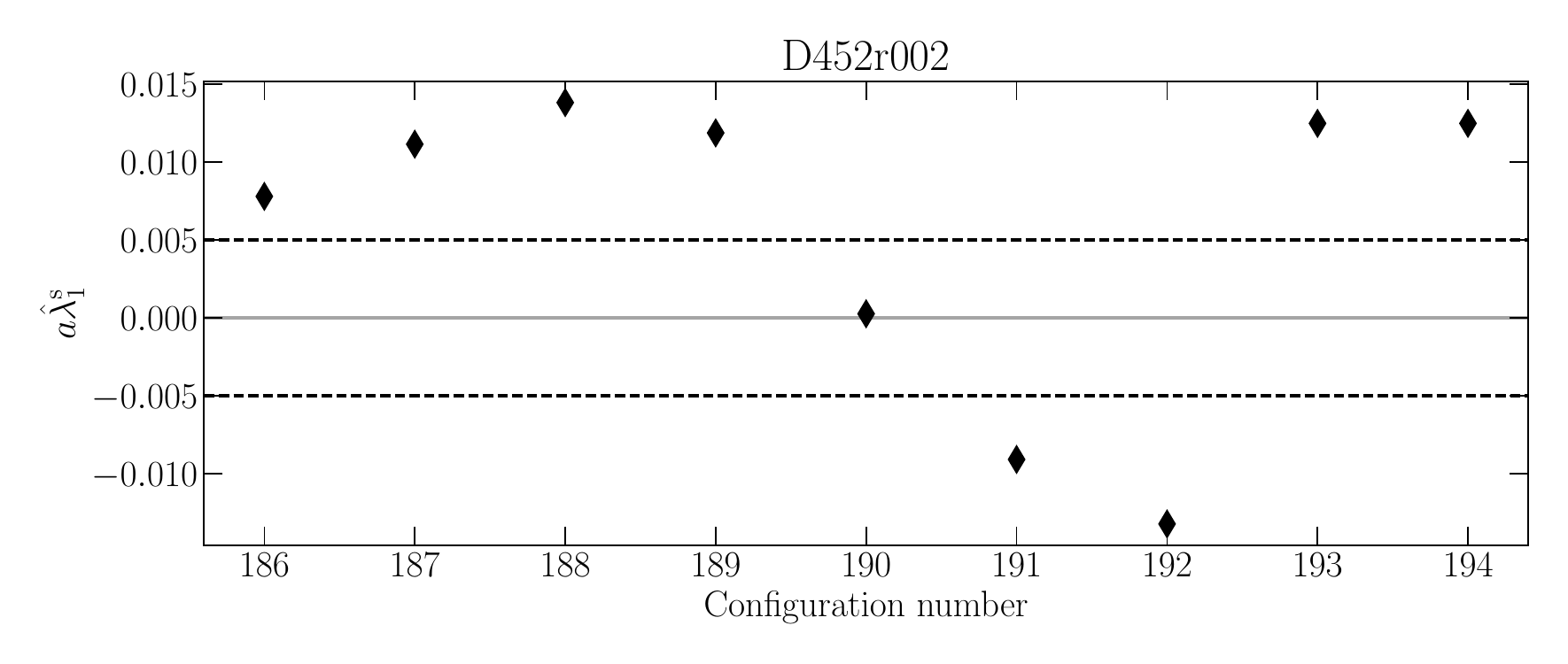}
	\caption{Part of the Monte Carlo history of the lowest eigenvalue of $\hat{Q}_{\rm s}$ with the sign according to \cite{Mohler:2020txx}. The assumed spectral gap $r_{\rm a}=0.005$ that was chosen for the simulation is depicted by dashed black lines. \label{f:D452r002_lambdasMC}}
\end{figure}

Smoothly fluctuating RHMC reweighting factors for the strange quark determinant
can be observed on the vast majority of the CLS ensembles, indicating a 
proper rational approximation during the simulation. 
On some of the ensembles, deviations towards smaller values of $W_{\rm s}$ can
be observed on a few configurations, indicating a lowest eigenvalue 
$\hat{\lambda}_1^{\rm s}$ below the parameter $r_{\rm a}$ that was chosen for 
the RHMC.

On a handful of configurations with an exceptionally low strange eigenvalue,
the convergence of the Taylor series in eq.~(\ref{e:RHMC_Z_Taylor}) is
particularly slow, so that $\mathrm{O}(100)$ elements are not sufficient
to achieve the required precision. 
An investigation of the lowest eigenvalue of $|\hat{Q}_{\rm s}|$ 
together with the information about its sign from \cite{Mohler:2020txx}
in the vicinity of a problematic configuration helps to understand the 
behavior of the simulation.
In Fig.~\ref{f:D452r002_lambdasMC} we show the Monte Carlo history of 
the lowest eigenvalue of $\hat{Q}_{\rm s}$ on the D452r002 ensemble around
configuration number 190, where convergence issues are found.
A departure in the region of negative
eigenvalues can be observed for configurations 191 and 192. When switching from
positive to negative eigenvalues, the exceptionally small eigenvalue
$2.6\cdot10^{-4}$, well below $r_{\rm a}^{\rm D452} = 5\cdot10^{-3}$, is encountered on configuration 190.%
\footnote{
Note that the ensembles studied in this work were generated
with a constant sum of the bare light and strange quark masses 
along a chiral trajectory \cite{Bietenholz:2010jr}. 
Therefore, the strange quark is lighter than physical for unphysically large
light quark masses, cf.\ table~\ref{t:enstab}.}

This is exactly the situation that has been the subject of discussion in 
section~\ref{s:RHMC_stoch}. Such a small eigenvalue poses two problems for 
the computation of the reweighting factor $W_{\rm s}$: 
A single random source may not be sufficient to estimate $W_{\rm s}$ with
precision, similarly to the case of the twisted-mass reweighting
factor described above. Secondly the estimation along the lines of section
\ref{s:RHMC_stoch} may fail because the Taylor series does not converge at all.

We have found that both issues are resolved when low-mode deflation 
using $N_{\rm L}=1$ is applied in the computation of $W_{\rm s}$, because
any significant deviation from the ensemble average is accounted for by 
the exactly evaluated contribution of the lowest eigenmode. 
This is the case for all investigated configurations where $W_{\rm s}$ could
not be determined at all or not be determined accurately from stochastic
estimation.

\section{Conclusions \label{s:conclusions}}
In the course of simulating lattice QCD with $\mathrm{O}(a)$ improved Wilson
fermions and twisted-mass reweighting, it has been found that small but
imprecisely determined twisted-mass reweighting factors can negatively affect
the precise estimation of ensemble averages, especially for observables
sensitive to the low eigenmodes of the light Dirac operator.%
\footnote{Thanks to the stochastic estimation of the reweighting factor in the traditional setup, the correctness of the results is not in question in this case.}%
We have shown that this issue can be resolved by exact deflation
of a small number of low modes in the computation of the corresponding 
reweighting factor. The procedure can be implemented without much effort and
may even reduce the overall cost of the computation. 

Two other improved estimators for the twisted-mass reweighting factor have
been introduced in the past. One is based on a frequency splitting of the
quark determinant, as described in section~\ref{s:theo_tm_stoch}, and requires
a much larger amount of computer time on all configurations of an 
ensemble to more accurately estimate a small number of reweighting factors. 
The resulting precision at tolerable cost is not competitive 
with the observed precision of deflated reweighting factors.
The strategy of \cite{RQCD:2022xux} relies on a larger number of stochastic
samples on configurations with small reweighting factors and a more robust
estimator for the stochastic estimator.
Low-mode deflation, on the other hand, allows to use a small number of noise sources for the estimation of the stochastic remainder on all configurations.

We have applied low-mode deflation to the computation of the 
twisted-mass reweighting factors on 15 CLS ensembles with 
pion masses between $360\,$MeV and $132\,$MeV at lattice spacings 
$a\geq 0.05\,$fm. The newly computed set of reweighting factors will 
be used in future work.
The impact on the overall precision of phenomenologically relevant 
observables at the physical point depends on the observable in question 
and its sensitivity to the low modes of the Dirac operator. 
We have discussed the impact on the computation of $af_\pi$ on a
particularly affected ensemble in section \ref{s:num_tm}. 
This directly influences the attainable precision in scale setting 
and the computation of the hadronic vacuum polarization contribution 
to the muon $g-2$.
Indications of stabilizing behavior on the computation have also been 
observed in the study of $I=1$ $\pi$-$\pi$ scattering at the 
physical point \cite{Paul:2021pjz}.

As pointed out in the introduction, the precise estimation of the reweighting
factor alone may not be sufficient to guarantee the cancellation of 
fluctuations in observables and reweighting factors to the amount that is
required for high-precision calculations.
Examples of such fluctuations are shown for the nucleon sigma term in
\cite{Agadjanov:2023jha}, where the reweighting factors of this work have been
employed, and for nucleon three-point functions in \cite{Bali:2023sdi}.
We note in passing that nucleon three-point functions are particularly
affected by exceptionally small eigenvalues because, when expressed in terms 
of eigenmodes, each propagator carries a factor $\hat{\lambda}_1^{-1}$ which 
may enhance fluctuations of the local magnitude of eigenmodes as the ones
described in \cite{Giusti:2003gf}.
Improved estimators for the correlation functions may be needed 
to remove these outliers completely.
Low-mode averaging \cite{Giusti:2004yp, DeGrand:2004qw, Giusti:2005sx} is 
expected to have a stabilizing effect because the contributions of the lowest
eigenmodes of $\hat{Q}$ can be exactly taken into account, as it is the case
for the deflated reweighting factor. 

While the strange quark Wilson-Dirac operator was long time assumed to have
a solid spectral gap, it was found that sign changes of the quark
determinant have a non-vanishing probability at finite lattice spacing \cite{Mohler:2020txx}. 
In the vicinity of some of these sign flips, problems in the computation 
of the strange quark RHMC reweighting factor occur when the lowest eigenvalue
of the Dirac operator is exceptionally small.
This leads to imprecise estimates or the failure of the computation. 
We have found that low-mode deflation can be applied here to overcome these
problems by exactly computing the contribution of the lowest eigenmode to the reweighting factor.

With the results of this work, in combination with the findings of
\cite{Mohler:2020txx}, the reweighting factors that are needed 
to restore the target distribution for modern simulations of Wilson quarks
based on twisted-mass reweighting and the RHMC algorithm can be computed with
high precision. This is the basis for precise predictions in the 
low energy regime of QCD.
\begin{acknowledgement}%
We like to thank Gunnar Bali, Dalibor Djukanovic, Fabian Joswig, Daniel Mohler,
Rainer Sommer, Simon Weishäupl and the members of the Mainz lattice group 
for discussions on the matter of computing reweighting factors.
We thank Hartmut Wittig for a critical reading of an earlier version of 
the manuscript.
Thanks go to Marco C\`e and Andreas Risch for providing the light quark 
correlation functions from stochastic sources on ensemble D150.
We are grateful to our colleagues in the CLS initiative for sharing ensembles.
Calculations for this project have been performed on the HPC clusters
HIMster-II at Helmholtz Institute Mainz and Mogon-II at Johannes Gutenberg-Universität (JGU) Mainz.
The estimation of statistical uncertainties of gauge averages in this work 
has been performed using the $\Gamma$-method in the implementation of 
the \texttt{pyerrors} package \cite{Wolff:2003sm,Ramos:2018vgu,Joswig:2022qfe}.
Plots have been generated with \texttt{matplotlib} \cite{Hunter:2007}.
\end{acknowledgement}

\appendix
\section{The spectral gap of the light quark Wilson-Dirac operator\label{a:gap}}
The distribution of the spectral gap of the Dirac-Wilson operator has been investigated in some detail in the two-flavor theory \cite{DelDebbio:2005qa, DelDebbio:2007pz}, where the first of the two references focused on the 
theory without $\mathrm{O}(a)$ improvement. Information on the behavior of exponentiated-clover Wilson fermions in the $2+1$ flavor theory is provided in \cite{Francis:2019muy}.\footnote{In all of those works, the gap of $Q$ has been investigated. In the following, we will work with $\hat{Q}$.}%
In this appendix, we will provide some insight in the distribution of the lowest eigenvalue $\hat{\lambda}_1$ of the even-odd preconditioned hermitian light-quark Wilson-clover Dirac operator $\hat{Q}_{\rm l}$ 
on the $2+1$ flavor CLS ensembles that have been included in this study.

We define spectral gap of $\hat{Q}_{\rm l}$ as as the absolute value of its smallest eigenvalue $\hat{\lambda}_1$.
Because the tail of the distribution of the spectral gap might be poorly
sampled along a Monte Carlo chain, we do not rely on the 
standard deviation to estimate the width of the distribution. 
Instead, the median and width $\hat{\sigma}$ 
of the reweighted eigenvalue distributions are computed as proposed 
in appendix D of \cite{Francis:2019muy}.

\begin{figure}[th]
	\centering
	\includegraphics[width=.9\textwidth]{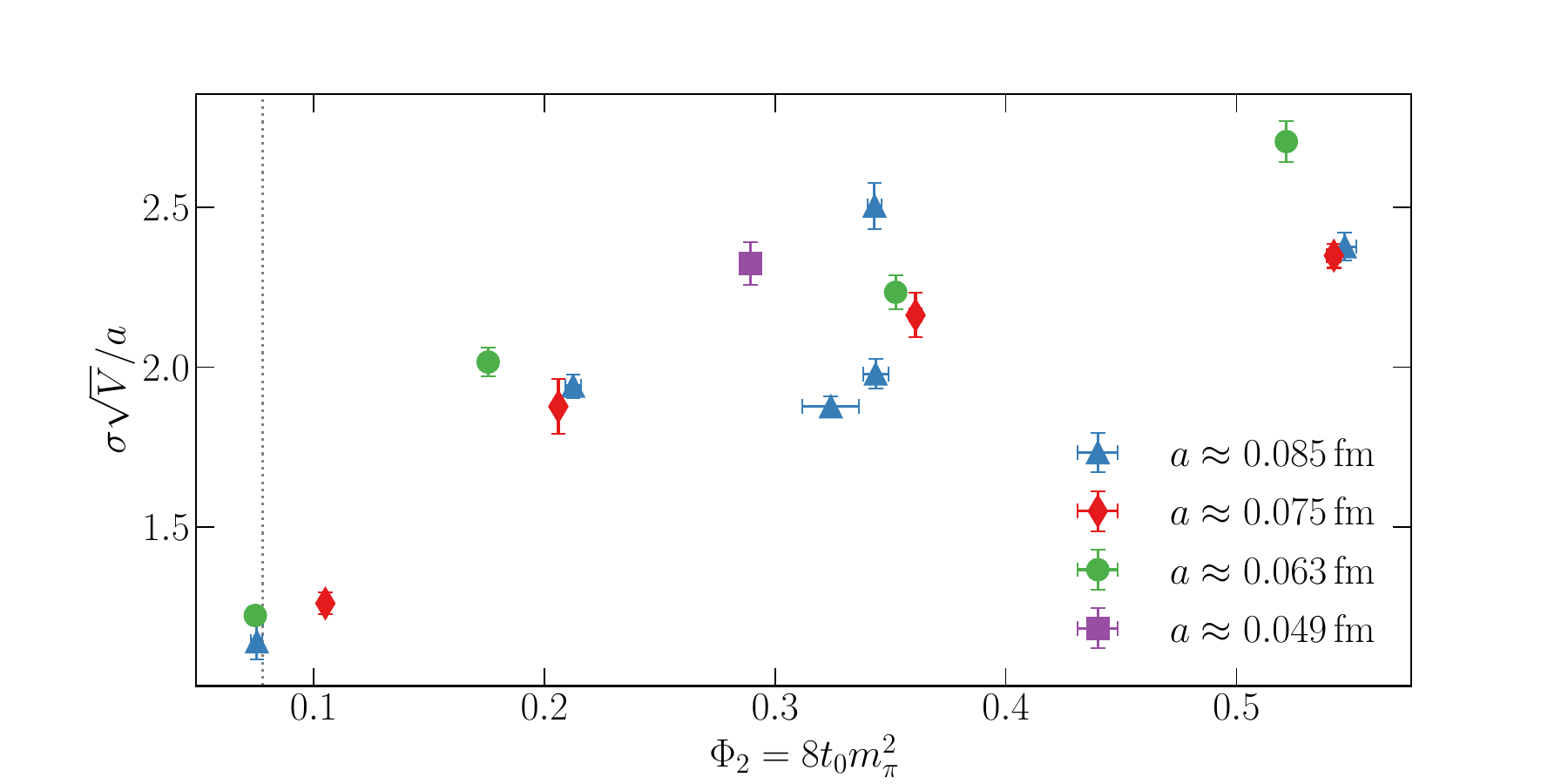}
	\caption{Width of the reweighted gap distributions, scaled by $\sqrt{V}/a$, versus $\Phi_2$ (from \cite{Ce:2022kxy}) as proxy for the light quark mass. The uncertainties of the widths have been estimated using a bootstrap procedure.
	The vertical line denotes physical light quark masses. \label{f:gap_width}}
\end{figure}

In Figure~\ref{f:gap_width} we plot the product of the width $\hat{\sigma}$ 
and the
square root of the four volume $V$ versus $\Phi_2= 8 t_0 m_\pi^2$. The latter 
may be used as proxy for the light quark mass to leading order in 
chiral perturbation theory \cite{Bruno:2016plf}.
In the two-flavor theory without $\mathrm{O}(a)$ improvement, the product
$\sigma\sqrt{V}/a$ was found to be approximately constant, where $\sigma$ was
the width of the distribution of the spectral gap of $Q$. In the improved
two-flavor theory \cite{DelDebbio:2007pz}, a dependence of 
$\sigma \sqrt{V}/a$ on the quark mass
was observed. This is also the case in our study, where $\hat{\sigma}\sqrt{V}/a$
seems to decrease towards smaller light quark masses. 

\begin{figure}[th]
	\centering	\includegraphics[width=.9\textwidth]{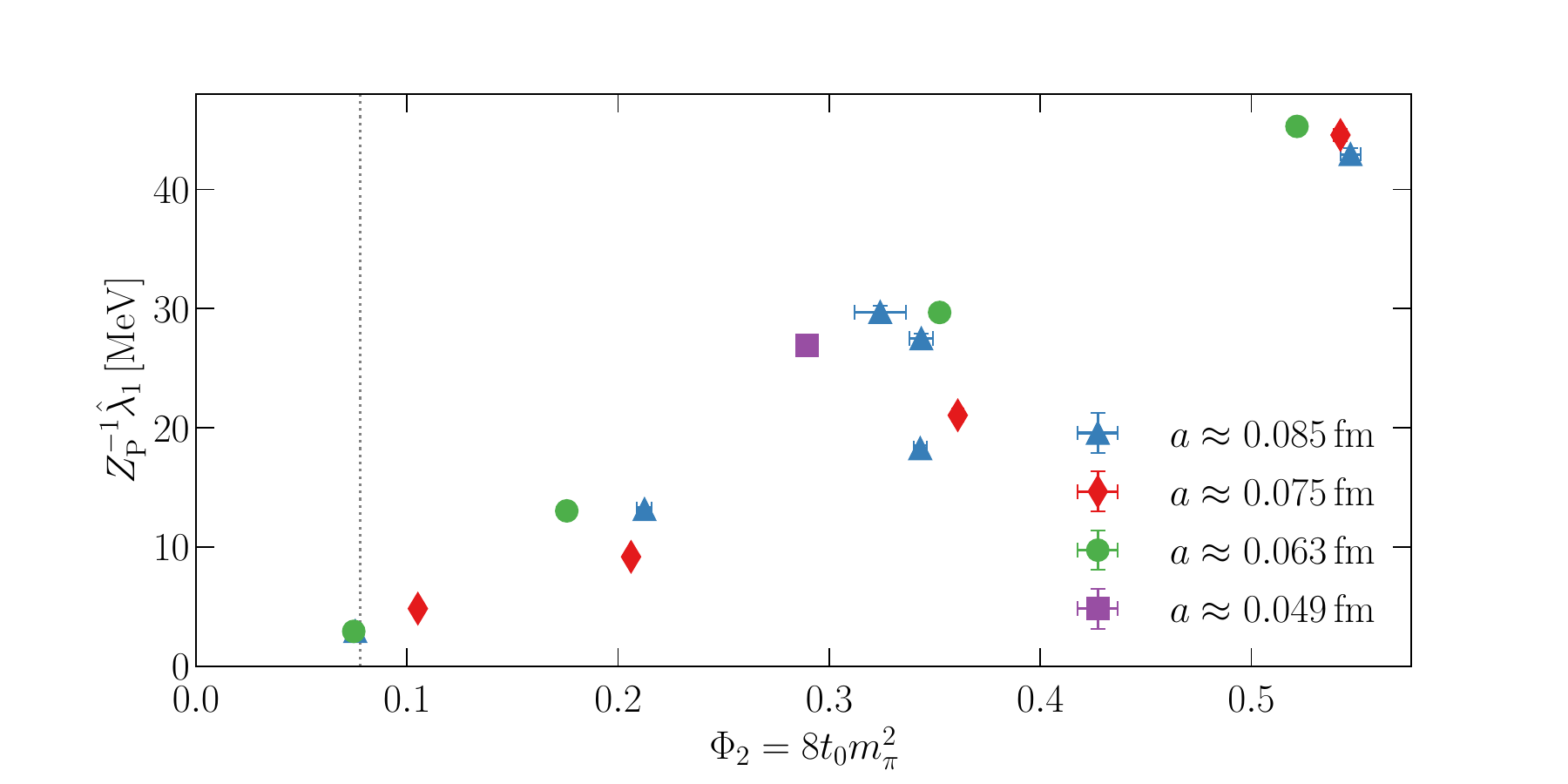}
	\caption{Reweighted spectral gap $\hat{\lambda}_1/Z_{\rm P}$ versus $\phi_2$. Note that the symbols of the two ensembles D150 and E250, 
	both approximately at physical light quark mass, overlap. For the 
	majority of data points, the error bars are smaller than the symbol 
	sizes. \label{f:gap_quark}}
\end{figure}

Figure~\ref{f:gap_quark} shows the dependence of the mean of the distribution
of $\langle \hat{\lambda}_1 \rangle$, renormalized with $1/Z_{\rm P}$ 
from \cite{Campos:2018ahf}, on the light quark proxy $\Phi_2$.
In formulations of lattice QCD that preserve chiral symmetry, the gap of $Q_{\rm l}$ is bounded from below by the bare current-quark mass $m_{12}$ 
\cite{DelDebbio:2005qa}. While we cannot expect this relation to hold for
Wilson quarks, the data in Fig.~\ref{f:gap_quark} seems to be approximately 
proportional to $\Phi_2$.

\begin{table}[th]
	\centering
	\begin{tabular}{cccccc}
		id & a\,[fm] & $m_\pi\,$[MeV] & $m_K\,$[MeV] & $L$\,[fm] & $Lm_\pi$ \\
\hline
H102 & 0.085 & 356 & 442 & 2.7 & 4.9 \\
U101 &       & 274 & 467 & 2.0 & 2.8 \\
H105 &       & 282 & 468 & 2.7 & 3.9 \\
N101 &       & 282 & 467 & 4.1 & 5.8 \\
C101 &       & 222 & 478 & 4.1 & 4.6 \\
D150 &       & 132 & 487 & 5.4 & 3.6 \\
\hline
S400 & 0.075 & 355 & 445 & 2.4 & 4.3 \\
N451 &       & 289 & 466 & 3.6 & 5.3 \\
D450 &       & 219 & 482 & 4.8 & 5.3 \\
D452 &       & 156 & 490 & 4.8 & 3.8 \\
\hline
N203 & 0.064 & 348 & 445 & 3.1 & 5.4 \\
N200 &       & 286 & 466 & 3.1 & 4.4 \\
D200 &       & 202 & 484 & 4.1 & 4.2 \\
E250 &       & 132 & 495 & 6.1 & 4.1 \\
\hline
J303 & 0.049 & 259 & 478 & 3.2 & 4.1 \\

	\end{tabular}
	\caption{Overview of ensembles used in this work. The lattice spacing $a$
		has been computed in \cite{RQCD:2022xux}. Pion and kaon masses have
		been determined in the context of \cite{Ce:2022kxy}. 
		$L$ denotes the spatial extent of the box. \label{t:enstab}}
\end{table}

\small
\addcontentsline{toc}{section}{References}
\bibliographystyle{JHEP}
\bibliography{library}

\end{document}